\documentclass[letterpaper,english,9pt]{scrartcl}
\usepackage[T1]{fontenc}
\usepackage[latin9]{inputenc}
\usepackage{mathrsfs}
\usepackage{amsmath}
\usepackage{amsthm}
\usepackage{amssymb}
\usepackage{graphicx}
\usepackage{esint}

\makeatletter

\theoremstyle{plain}
\newtheorem{thm}{\protect\theoremname}

\usepackage{spconf}

\usepackage{dsfont}\usepackage{babel}
\usepackage{relsize}\usepackage{subdepth}

\name{Dionysios S. Kalogerias and Athina P. Petropulu\thanks{This work was supported by the National Science Foundation (NSF) under
Grant CNS-1239188.}}
\address{Department of Electrical \& Computer Engineering \\Rutgers, The State University of New Jersey, Piscataway, NJ 08854, USA}

\usepackage{babel}

\providecommand{\theoremname}{Theorem}

\usepackage{babel}

\providecommand{\theoremname}{Theorem}

\makeatother

\usepackage{babel}
\providecommand{\theoremname}{Theorem}

\begin{document}

\title{Mobile Beamforming \& Spatially Controlled Relay Communications}
\maketitle
\begin{abstract}
We consider stochastic motion planning in single-source single-destination
robotic relay networks, under a cooperative beamforming framework.
Assuming that the communication medium constitutes a spatiotemporal
stochastic field, we propose a $2$-stage stochastic programming formulation
of the problem of specifying the positions of the relays, such that
the \textit{expected} \textit{reciprocal} of their total beamforming
power is maximized. Stochastic decision making is made on the basis
of random causal CSI. Recognizing the intractability of the original
problem, we propose a lower bound relaxation, resulting to a nontrivial
optimization problem with respect to the relay locations, which is
equivalent to a small set of simple, tractable subproblems. Our formulation
results in spatial controllers with a predictive character; at each
time slot, the new relay positions should be such that the expected
power reciprocal at the next time slot is maximized. Quite interestingly,
the optimal control policy to the relaxed problem is purely selective;
under a certain sense, \textit{only the best relay should move}.
\end{abstract}

\vspace{-0.2cm}

\section{Introduction}

\vspace{-0.1cm}

Cooperative beamforming constitutes a powerful method for information
relaying in multi-hop networks. It is well known to greatly improve
communication reliability by increasing directional channel gain,
enabling low power, long distance transmissions with fewer hops, and
with minimal interference \cite{Havary_BEAM_2008,Beamforming_1_2009,Beamforming_2_2009}.
In Amplify-and-Forward (AF) beamforming, which is considered here
due to its simplicity \cite{Havary_BEAM_2008}, typically, the objective
is to determine source power and/or relay beamforming weights so that
certain optimality criteria are met, such as Quality-of-Service (QoS)
at the destinations, or transmit power at the relays \cite{Havary_BEAM_2008,Beamforming_1_2009,Beamforming_2_2009}.
This optimization procedure critically depends on availability of
Channel State Information (CSI) at the relays. In the literature,
CSI is mostly assumed either known \cite{Beamforming_1_2009,Beamforming_3_Petropulu2009,Havary_BEAM_2010},
or unknown with known statistics \cite{Havary_BEAM_2008,Beamforming_4_2008,Beamforming_5_2009,Beamforming_6_Petropulu2011,Beamforming_7_Petropulu2011,Beamforming_8_Petropulu2012},
with the latter assumption better reflecting reality.

Recently, there has been works which exploit mobility of the relays
assisting the communication, in order to further enhance performance
in beamforming networks. In \cite{NikosBeam-2}, mobility control
has been combined with optimal transmit beamforming for transmit power
minimization, while maintaining QoS in multiuser cooperative networks.
Also, more recently \cite{KalPet-Jammers-2013}, under a slightly
different formulation, in the context of information theoretic secrecy,
decentralized mobility control has been jointly combined with noise
nulling and cooperative jamming for secrecy rate maximization in mobile
jammer assisted cooperative communication networks with one source,
one destination and multiple jammers. In the above works, the communication
channels among the entities of the network (or the related second
order statistics) have been assumed to be fixed during the whole motion
of the relays. However, this assumption might be restrictive in scenarios
where the channels change dynamically/stochastically through time
and space.

In this paper, we present a novel treatment of the basic AF single-source
single-destination relay beamforming problem, under the fundamental
assumption that the channel, on the basis of which control decisions
are determined, constitutes a \textit{spatiotemporal stochastic process}.
More specifically, we consider a time slotted, spatially controlled
communication system, where, at each time slot, both communications
and relay motion take place. Under this model, we propose a $2$-\textit{stage
stochastic programming} formulation of the problem of specifying the
positions of the relays, such that the \textit{reciprocal} of their
total beamforming power is maximized \textit{on average}, based on
causal CSI. The proposed formulation results in relay spatial controllers
with a predictive character; at each time slot, the decision on the
new positions should be such that the expected power reciprocal, occurring
\textit{at the next time slot}, is maximized. Due to the intractability
of the original problem, we propose a lower bound relaxation, which
provably results to a nontrivial optimization problem with respect
to the positions of the relays. Under a realistic ``log-normal''
wireless channel model \cite{MostofiSpatial2012}, the relaxed problem
is equivalent to solving a set of two dimensional, computationally
tractable subproblems. Quite remarkably, the optimal control policy
to the relaxed problem is purely selective; under a certain sense,
\textit{only the best relay should move}. 

\vspace{-0.2cm}

\section{System Model}

\vspace{-0.1cm}

On a closed planar region ${\cal S}\subset\mathbb{R}^{2}$, we consider
a wireless cooperative network consisting of one source, one destination
and $R\in\mathbb{N}^{+}$ assistive relays. Each entity of the network
is equipped with a single antenna, being able for both information
reception and broadcasting/transmission. The source and destination
are stationary and located at ${\bf p}_{S}\in{\cal S}$ and ${\bf p}_{D}\in{\cal S}$,
respectively, whereas the relays are assumed to be mobile; each relay
$i\in\mathbb{N}_{R}^{+}$ moves along a trajectory ${\bf p}_{i}\left(t\right)\in{\cal S}$,
where, in general, $t\in\mathbb{R}_{+}$. We also define the supervector
${\bf p}\left(t\right)\triangleq\left[{\bf p}_{1}^{\boldsymbol{T}}\left(t\right)\,{\bf p}_{2}^{\boldsymbol{T}}\left(t\right)\,\ldots\,{\bf p}_{R}^{\boldsymbol{T}}\left(t\right)\right]^{\boldsymbol{T}}\in{\cal S}^{R}$.
Additionally, we assume that the relays can cooperate with each other,
either in terms of local message exchange, or by communicating with
a local fusion center, through a dedicated channel.

Assuming that a direct link between the source and the destination
does not exist, the role of the relays is determined to be assistive
to the communication, in a classical two phase AF sense. Fix a $T>0$,
and \textit{divide the time interval $\left[0,T\right]$ into $N_{T}$
time slots}, with $t\in\mathbb{N}_{N_{T}}^{+}$ denoting the respective
time slot. Let $s\left(t\right)\in\mathbb{C}$, with $\mathbb{E}\left\{ \left|s\left(t\right)\right|^{2}\right\} \equiv1$,
denote the symbol to be transmitted at time slot $t$. Also, assuming
a flat fading channel model, as well as channel reciprocity and quasistaticity
in each time slot, let the sets $\left\{ f_{i}\left({\bf p}_{i}\left(t\right),t\right)\in\mathbb{C}\right\} _{i\in\mathbb{N}_{R}^{+}}$
and $\left\{ g_{i}\left({\bf p}_{i}\left(t\right),t\right)\in\mathbb{C}\right\} _{i\in\mathbb{N}_{R}^{+}}$
contain the \textit{random, spatiotemporally varying} source-relay
and relay-destination channel gains, respectively. Then, if $P_{0}>0$
denotes the transmission power, during AF phase $1$, relay $i$ receives
the amplified symbol $\sqrt{P_{0}}s\left(t\right)$, modulated by
$f_{i}\left({\bf p}_{i}\left(t\right),t\right)$, plus an additive,
spatiotemporally white noise component $n_{i}\left(t\right)\in\mathbb{C}$,
with $\mathbb{E}\left\{ \left|n_{i}\left(t\right)\right|^{2}\right\} \equiv\sigma^{2}$,
for all $i\in\mathbb{N}_{R}^{+}$. During AF phase $2$, all relays
simultaneously retransmit the information received, each modulating
their received signal by a weight $w_{i}\left(t\right)\in\mathbb{C},i\in\mathbb{N}_{R}^{+}$.
The signal received at the destination can be expressed as the superposition
of the weighted relay signals, plus another spatiotemporally white
noise component $n_{D}\left(t\right)\in\mathbb{C}$, with $\mathbb{E}\left\{ \left|n_{D}\left(t\right)\right|^{2}\right\} \equiv\sigma_{D}^{2}$.

In the following, whereas it is assumed that the stochastic processes
$f_{i}\left({\bf p}_{i}\left(t\right),t\right)$ and $g_{j}\left({\bf p}_{t}\left(t\right),t\right)$
may be \textit{statistically dependent both spatially and temporally},
for all $\left(i,j\right)\in\mathbb{N}_{R}^{+}\times\mathbb{N}_{R}^{+}$,
it is also assumed that, \textit{as usual}, the random processes $s\left(t\right)$,
$\left[\left\{ f_{i}\left({\bf p}_{i}\left(t\right),t\right),g_{i}\left({\bf p}_{i}\left(t\right),t\right)\right\} _{i\in\mathbb{N}_{R}^{+}}\right]$,
$n_{i}\left(t\right)$ for all $i\in\mathbb{N}_{R}^{+}$, and $n_{D}\left(t\right)$
are mutually independent. Lastly, we will assume that, at each time
slot $t,$ CSI $\left\{ f_{i}\left({\bf p}_{i}\left(t\right),t\right)\right\} _{i\in\mathbb{N}_{R}^{+}}$
and $\left\{ g_{i}\left({\bf p}_{i}\left(t\right),t\right)\right\} _{i\in\mathbb{N}_{R}^{+}}$
is known \textit{exactly} to all relays. This may be achieved through
pilot based estimation and will be considered a valid practical assumption.

\vspace{-0.2cm}

\section{Wireless Channel Modeling}

\vspace{-0.1cm}

At each time slot $t\in\mathbb{N}_{N_{T}}^{+}$, the $i$-th source-relay
channel gain can be decomposed as \cite{Goldsmith2005Wireless}
\begin{flalign}
f_{i}\left({\bf p}_{i}\left(t\right)\hspace{-2pt},t\right) & \equiv\underbrace{f^{PL}\hspace{-2pt}\left({\bf p}_{i}\left(t\right)\right)}_{\text{path loss}}\underbrace{f_{i}^{SH}\hspace{-2pt}\left(t\right)}_{\text{shadowing}}\underbrace{f_{i}^{MF}\hspace{-2pt}\left(t\right)}_{\text{fading}}e^{\mathfrak{J}\frac{2\pi d_{iS}\left(t\right)}{\lambda}},\label{eq:Channel_1}
\end{flalign}
where $\mathfrak{J}\triangleq\sqrt{-1}$, $\lambda>0$ denotes the
wavelength employed for the communication, and where: \textbf{1) }$f^{PL}\left({\bf p}_{i}\left(t\right)\right)\triangleq\left\Vert {\bf p}_{i}\left(t\right)-{\bf p}_{S}\right\Vert _{2}^{-\ell/2}$
$\triangleq\left(d_{iS}\left(t\right)\right)^{-\ell/2}$, where $\ell>0$
denotes the path loss exponent.\textbf{}\linebreak{}
\textbf{2)} $f_{i}^{SH}\left(t\right)\in\mathbb{R}$ denotes the shadowing
part of the channel, whose square is a base-$10$ log-normal random
variable with zero location. \textbf{3) }$f_{i}^{MF}\left(t\right)\in\mathbb{C}$
represents multipath fading, which is assumed to be an unpredictable,
\textit{spatiotemporally} white \cite{MostofiSpatial2012}, strictly
stationary process with known statistics. In particular, its phase,
$\phi_{f_{i}}\left(t\right)$, is assumed to be a white noise process
\textit{uniformly distributed} in $\left[-\pi,\pi\right]$, also independent
of its magnitude.

Now, since the complex exponential in \eqref{eq:Channel_1} is known,
let us substitute $f_{i}\left({\bf p}_{i}\left(t\right),t\right)\leftarrow\exp\left(-\mathfrak{J}2\pi d_{iS}\left(t\right)/\lambda\right)$$f_{i}\left({\bf p}_{i}\left(t\right),t\right)$.
Instead of working with the multiplicative model described by \eqref{eq:Channel_1},
it is much preferable to work in logarithmic scale. We may then define
\begin{align}
F_{i}\left({\bf p}_{i}\left(t\right),t\right) & \triangleq\alpha_{S}^{i}\left({\bf p}_{i}\left(t\right)\right)\ell+\sigma_{S}^{i}\left(t\right)+\xi_{S}^{i}\left(t\right),\;\forall i\in\mathbb{N}_{R}^{+}\label{eq:Amplitude_log}
\end{align}
and $\forall t\in\mathbb{N}_{N_{T}}^{+}$, where $\alpha_{S}^{i}\left({\bf p}_{i}\left(t\right)\right)\triangleq-10\log_{10}\left(d_{iS}\left(t\right)\right)$,
$\sigma_{S}^{i}\left(t\right)\triangleq10\log_{10}\left(f_{i}^{SH}\left(t\right)\right)^{2}$
and $\xi_{S}^{i}\left(t\right)\triangleq\overline{10\log_{10}\left|f_{i}^{MF}\left(t\right)\right|^{2}}$,
with $\overline{\left(\cdot\right)}$ denoting the zero mean version
of a random variable. Of course, we may stack all the $F_{i}\left({\bf p}_{i}\left(t\right),t\right)$'s
defined in \eqref{eq:Amplitude_log}, resulting in the vector additive
model
\begin{equation}
\boldsymbol{F}\left({\bf p}\left(t\right),t\right)\triangleq\boldsymbol{\alpha}_{S}\left({\bf p}\left(t\right)\right)\ell+\boldsymbol{\sigma}_{S}\left(t\right)+\boldsymbol{\xi}_{S}\left(t\right)\in\mathbb{R}^{R\times1},\label{eq:Vector_1}
\end{equation}
where $\boldsymbol{\alpha}_{S}\left(t\right)$, $\boldsymbol{\sigma}_{S}\left(t\right)$
and $\boldsymbol{\xi}_{S}\left(t\right)$ are defined accordingly.
We can also define $\boldsymbol{G}\left({\bf p}\left(t\right),t\right)\triangleq\boldsymbol{\alpha}_{D}\left({\bf p}\left(t\right)\right)\ell+\boldsymbol{\sigma}_{D}\left(t\right)+\boldsymbol{\xi}_{D}\left(t\right)\in\mathbb{R}^{R\times1},$
with each quantity in direct correspondence with \eqref{eq:Vector_1}.

The spatiotemporal dynamics of $\left\{ f_{i}\left({\bf p}_{i}\left(t\right),t\right)\right\} _{i}$
and\linebreak{}
$\left\{ g_{i}\left({\bf p}_{i}\left(t\right),t\right)\right\} _{i}$
are modeled through those of the shadowing components of $\left\{ F_{i}\left({\bf p}_{i}\left(t\right),t\right)\right\} _{i}$
and $\left\{ G_{i}\left({\bf p}_{i}\left(t\right),t\right)\right\} _{i}$.
It is assumed that for any $N_{T}$, the process $\left[\left\{ \boldsymbol{F}^{\boldsymbol{T}}\left({\bf p}\left(t\right),t\right)\,\boldsymbol{G}^{\boldsymbol{T}}\left({\bf p}\left(t\right),t\right)\right\} _{t\in\mathbb{N}_{N_{T}}^{+}}\right]^{\boldsymbol{T}}$
is jointly Gaussian with known means and known covariance matrix.
More specifically \cite{MostofiSpatial2012}, $\xi_{D\left(S\right)}^{i}\left(t\right)\overset{i.i.d.}{\sim}{\cal N}\left(0,\sigma_{\xi}^{2}\right)$,
for all $t\in\mathbb{N}_{N_{T}}^{+}$ and $i\in\mathbb{N}_{R}^{+}$
\cite{Cotton2007}. Second, extending Gudmundson's model \cite{Gudmundson1991}
in a straightforward way, we propose defining the spatiotemporal correlations
of the shadowing part of the channel as
\begin{align}
\mathbb{E}\left\{ \sigma_{S}^{i}\left(k\right)\sigma_{S}^{j}\left(l\right)\right\}  & \triangleq\eta^{2}e^{-\frac{\left\Vert {\bf p}_{i}\left(k\right)-{\bf p}_{j}\left(l\right)\right\Vert _{2}}{\beta}-\frac{\left|k-l\right|}{\gamma}},
\end{align}
and correspondingly for $\left\{ \sigma_{D}^{i}\left(t\right)\right\} _{i\in\mathbb{N}_{R}^{+}}$,
and additionally,
\begin{align}
\mathbb{E}\left\{ \sigma_{S}^{i}\left(k\right)\sigma_{D}^{j}\left(l\right)\right\}  & \triangleq\mathbb{E}\left\{ \sigma_{S}^{i}\left(k\right)\sigma_{S}^{j}\left(l\right)\right\} e^{-\frac{\left\Vert {\bf p}_{S}-{\bf p}_{D}\right\Vert _{2}}{\delta}},
\end{align}
for all $\left(i,j\right)\in\mathbb{N}_{R}^{+}\times\mathbb{N}_{R}^{+}$
and all $\left(k,l\right)\in\mathbb{N}_{N_{T}}^{+}\times\mathbb{N}_{N_{T}}^{+}$.
In the above, $\eta^{2}$ and $\beta>0$ are called the \textit{shadowing
power} and the \textit{correlation distance}, respectively. In this
fashion, we will call $\gamma>0$ and $\delta>0$ the \textit{correlation
time} and the\textit{ BS (Base Station) correlation}, respectively.
For later reference, let us define the (cross)covariance matrices
$\boldsymbol{\Sigma}_{SD}\left(t_{k},t_{l}\right)\triangleq\mathbb{E}\left\{ \boldsymbol{\sigma}_{S}\left(t_{k}\right)\boldsymbol{\sigma}_{D}^{\boldsymbol{T}}\left(t_{l}\right)\right\} +\mathds{1}_{\left\{ S\equiv D\right\} }\mathds{1}_{\left\{ t_{k}\equiv t_{l}\right\} }\sigma_{\xi}^{2}{\bf I}_{R}$\linebreak{}
$\in\mathbb{S}_{+\left(+\right)}^{R}$, as well as $\boldsymbol{\Sigma}\left(t_{k},t_{l}\right)\triangleq\left[\boldsymbol{\Sigma}_{SS}\left(t_{k},t_{l}\right)\,\boldsymbol{\Sigma}_{SD}\left(t_{k},t_{l}\right);\right.$\linebreak{}
$\left.\boldsymbol{\Sigma}_{SD}\left(t_{k},t_{l}\right)\,\boldsymbol{\Sigma}_{DD}\left(t_{k},t_{l}\right)\right]\in\mathbb{S}_{+}^{2R}$.

\vspace{-0.2cm}

\section{Mobile Beamforming}

\vspace{-0.1cm}

At each time slot $t\in\mathbb{N}_{T}^{+}$ and assuming the same
carrier for all communication tasks, we employ a basic joint communication/decision
making TDMA-like protocol, as follows: \textbf{1)} The source broadcasts
a pilot signal to the relays, which then estimate the channels relative
to the source.\textbf{ 2)} The same procedure is carried out for the
channels relative to the destination.\textbf{ 3)} Then, based on the
estimated CSI, beamforming is implemented. \textbf{4)} Finally, based
on the CSI received \textit{so far}, the spatial controllers of the
relays are determined, implementing accurate stochastic decision making.

In the following, let $\left\{ \mathscr{C}\left({\cal T}_{t}\right)\right\} _{t\in\mathbb{N}_{N_{T}}^{+}}$
denote the set of channel gains observed by the relays, \textit{along
the path of their point trajectories} ${\cal T}_{i}\triangleq\left\{ {\bf p}\left(t\right)\right\} _{t\in\mathbb{N}_{i}^{+}},i\in\mathbb{N}_{N_{T}}^{+}$,
where ${\cal T}_{t}\equiv\left\{ {\cal T}_{t-1},{\bf p}\left(t\right)\right\} $.
Further, it is assumed that the motion of the relays obeys the differential
equation $\dot{{\bf p}}\left(\tau\right)\equiv{\bf u}\left(\tau\right)$,
for all $\tau\in\mathbb{R}_{+}$. Apparently, relay motion is in continuous
time. However, assuming the relays move \textit{only after their controls
have been determined and up to the start of the next time slot}, we
can write
\begin{equation}
{\bf p}\left(t\right)\equiv{\bf p}\left(t-1\right)+\int_{\Delta\tau_{t-1}}{\bf u}_{t-1}\left(\tau\right)\text{d}\tau,\quad\forall t\in\mathbb{N}_{N_{T}}^{+},\label{eq:motion_model_2}
\end{equation}
with ${\bf p}\left(1\right)\equiv{\bf p}{}_{init}$, and where $\Delta\tau_{t}\in\mathbb{R}$
denotes the time interval that the relays are allowed to move in each
time slot $t\in\mathbb{N}_{N_{T}}^{+}$. Of course, at each time slot
$t$, $\Delta\tau_{t}$ must be sufficiently small such that the temporal
correlations of the CSI at adjacent time slots are sufficiently strong.
These correlations are controlled by the correlation time parameter
$\delta$, which can be a function of the slot width. Therefore, the
velocity of the relays must be of the order of $\left(\Delta\tau_{t}\right)^{-1}$.
In this work, though, we assume that the relays are not resource constrained,
in terms of their robotic operation. See Fig. \ref{fig:Proposed_System}
for a block representation of the proposed joint beamforming and relay
motion control schema, where ${\cal I}_{t}$ contains the available
CSI at time slot $t$ and $\left\{ \cdot,\cdot\right\} $ denotes
the concatenation operation.

\vspace{-0.2cm}

\subsection{$2$-Stage Stochastic Optimization of Beamforming Weights \& Relay
Positions}

Suppose that, at time slot $t-1$, an oracle reveals $\mathscr{C}\left({\cal T}_{t}\right)$,
which of course includes the channels corresponding to the new positions
of the relays at the next time slot $t$. Then, \textit{given} $\mathscr{C}\left({\cal T}_{t}\right)$,
we are interested in determining $\boldsymbol{w}\left(t\right)\triangleq\left[w_{1}\left(t\right)\,w_{2}\left(t\right)\,\ldots\,w_{R}\left(t\right)\right]^{\boldsymbol{T}}$,
as the solution of the \textit{power reciprocal maximization} program
\begin{equation}
\begin{array}{rl}
\underset{{\bf w}\left(t\right)}{\mathrm{maximize}} & \hspace{3.07pt}\left(\mathbb{E}\left\{ \left.P_{R}\right|\mathscr{C}\left({\cal T}_{t}\right)\right\} \right)^{-1}\\
\mathrm{subject\,to} & \dfrac{\mathbb{E}\left\{ \left.P_{S}\right|\mathscr{C}\left({\cal T}_{t}\right)\right\} }{\mathbb{E}\left\{ \left.P_{I+N}\right|\mathscr{C}\left({\cal T}_{t}\right)\right\} }\ge\zeta
\end{array},\label{eq:Beamforming}
\end{equation}
where $P_{R}$, $P_{S}$ and $P_{I+N}$ denote the instantaneous power
at the relays, that of the signal component and that of the interference
plus noise component at the destination, and where $\zeta>0$ is chosen
such that \eqref{eq:Beamforming} is feasible. Note that instead of
minimizing the power at the relays, we are interested in maximizing
its inverse, as this facilitates the formulation of our joint communication-control
problem, as follows. Using the respective mutual independence assumptions,
\eqref{eq:Beamforming} can be written analytically and equivalently
as \cite{Havary_BEAM_2008}
\begin{equation}
\begin{array}{rl}
\underset{{\bf w}\left(t\right)}{\mathrm{maximize}} & \hspace{7.3pt}\left(\boldsymbol{w}^{\boldsymbol{H}}\left(t\right){\bf D}\left({\bf p}\left(t\right),t\right)\boldsymbol{w}\left(t\right)\right)^{-1}\\
\mathrm{subject\,to} & \dfrac{\boldsymbol{w}^{\boldsymbol{H}}\left(t\right){\bf R}\left({\bf p}\left(t\right),t\right)\boldsymbol{w}\left(t\right)}{\sigma_{D}^{2}+\boldsymbol{w}^{\boldsymbol{H}}\left(t\right){\bf Q}\left({\bf p}\left(t\right),t\right)\boldsymbol{w}\left(t\right)}\ge\zeta
\end{array},\label{eq:Beamforming_2}
\end{equation}
where, dropping the dependence on $\left({\bf p}\left(t\right),t\right)$
for brevity, 
\begin{flalign}
{\bf D} & \triangleq P_{0}\text{diag}\left(\left[\left|f_{1}\right|^{2}\,\left|f_{2}\right|^{2}\,\ldots\,\left|f_{R}\right|^{2}\right]^{\boldsymbol{T}}\right)+\sigma^{2}{\bf I}_{R}\in\mathbb{S}_{++}^{R},\\
{\bf R} & \triangleq P_{0}{\bf h}{\bf h}^{\boldsymbol{H}}\in\mathbb{S}_{+}^{R},\text{ with }{\bf h}\triangleq\left[f_{1}g_{1}\,f_{2}g_{2}\,\ldots\,f_{R}g_{R}\right]^{\boldsymbol{T}}\text{ and}\\
{\bf Q} & \triangleq\sigma^{2}\text{diag}\left(\left[\left|g_{1}\right|^{2}\,\left|g_{2}\right|^{2}\,\ldots\,\left|g_{R}\right|^{2}\right]^{\boldsymbol{T}}\right)\in\mathbb{S}_{++}^{R}.
\end{flalign}
Obviously, if the oracle could reveal $\mathscr{C}\left(\left\{ {\cal T}_{t-1},{\bf p}\left(t\right)\right\} \right)$
at $t-1$, we could further optimize the \textit{optimal value} of
\eqref{eq:Beamforming_2} with respect to ${\bf p}\left(t\right)$,
representing the new position of the relays. In the absence of the
oracle, though, this is impossible, since given $\mathscr{C}\left(\left\{ {\cal T}_{t-1}\right\} \right)$,
the channels at any position of the relays are nontrivial random variables.
However, given $\mathscr{C}\left(\left\{ {\cal T}_{t-1}\right\} \right)$,
it is reasonable to search for the best decision on the positions
of the relays at time slot $t$ (as a functional of $\mathscr{C}\left(\left\{ {\cal T}_{t-1}\right\} \right)$),
such that the optimal value of \eqref{eq:Beamforming_2} is maximized,
\textit{on average}. This results in the \textit{$2$-stage stochastic
program} \cite{Shapiro2009STOCH_PROG} 
\begin{equation}
\begin{array}{rl}
\underset{{\bf p}\left(t\right)}{\mathrm{maximize}} & \mathbb{E}\left\{ V\left({\bf p}\left(t\right),t\right)\right\} \\
\mathrm{subject\,to} & {\cal C}\left({\bf p}\left(t\hspace{-2pt}-\hspace{-2pt}1\right)\right)\ni{\bf p}\left(t\right)\equiv\mathbb{E}\left\{ \hspace{-2pt}\left.{\bf p}\left(t\right)\right|\mathscr{C}\left({\cal T}_{t-1}\right)\right\} 
\end{array},\label{eq:2STAGE-1}
\end{equation}
where
\begin{figure}
\centering\includegraphics{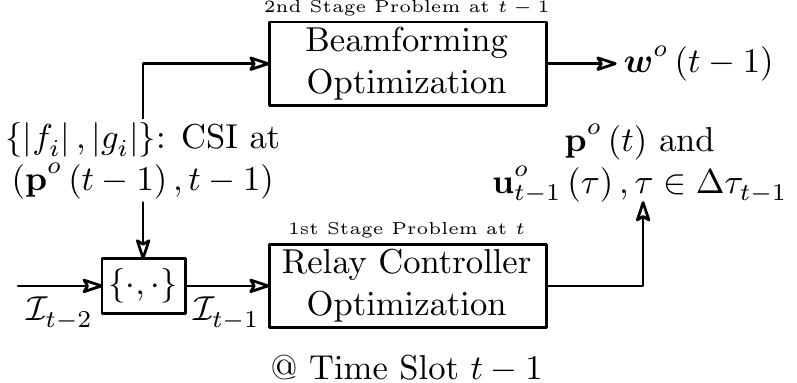}

\caption{\label{fig:Proposed_System}2-Stage optimization of beamforming weights
and spatial relay controllers. $\boldsymbol{w}^{o}\left(t-1\right)$,
${\bf u}_{t-1}^{o}\left(\tau\right)$ and ${\bf p}^{o}\left(t\right)$
denote the optimal beamforming weights and relay controllers at time
slot $t-1$ and the optimal relay positions at time slot $t$, respectively.
Also, ${\cal I}_{0}\equiv\left\{ \varnothing\right\} $.}
\end{figure}
 the random variable $V\left({\bf p}\left(t\right),t\right)$ (a functional
of $\mathscr{C}\left({\cal T}_{t}\right)$) denotes the optimal value
of \eqref{eq:Beamforming_2}, and ${\cal C}\left({\bf p}\left(t-1\right)\right)$
denotes a convex set, representing a spatially feasible neighborhood
around ${\bf p}\left(t-1\right)$. Problems \eqref{eq:2STAGE-1} and
\eqref{eq:Beamforming_2} are referred to as the \textit{first-stage
problem} and the \textit{second-stage problem}, respectively \cite{Shapiro2009STOCH_PROG}
(also see Fig. \ref{fig:Proposed_System}). In general, the existence
of elegant solutions for $2$-stage problems is extremely rare. Fortunately,
however, in our case, the second-stage problem admits a (semi)closed
form solution, expressed as \cite{Havary_BEAM_2008}
\begin{equation}
V=\dfrac{\lambda_{max}\left({\bf D}^{-1/2}\left({\bf R}-\zeta{\bf Q}\right){\bf D}^{-1/2}\right)}{\zeta\sigma_{D}^{2}}\triangleq\dfrac{\lambda_{max}\left({\bf B}\right)}{\zeta\sigma_{D}^{2}}.
\end{equation}
The resulting equivalent form for problem \eqref{eq:2STAGE-1} is
still difficult to solve because of two reasons; its \textit{variational
character} and, second, the fact that the expectation in the objective
cannot be computed in any reasonable and tractable way. By an application
of a generalized form of the \textit{fundamental lemma of stochastic
control} \cite{Speyer2008STOCHASTIC,Astrom1970CONTROL}, the first
stage problem \eqref{eq:2STAGE-1} can be equivalently replaced by
the much simpler problem
\begin{equation}
\underset{{\bf p}\left(t\right)\in{\cal C}\left({\bf p}\left(t-1\right)\right)}{\mathrm{maximize}}\mathbb{E}\left\{ \left.\lambda_{max}\left({\bf B}\left({\bf p}\left(t\right),t\right)\right)\right|\mathscr{C}\left({\cal T}_{t-1}\right)\right\} ,\label{eq:2STAGE-3}
\end{equation}
perfectly solving the first issue mentioned above. Regarding now the
second issue arising as a result of the (conditional) expectation
appearing in \eqref{eq:2STAGE-3}, due to the convexity of the maximum
eigenvalue operator in ${\bf B}$, we can use Jensen's Inequality
in order to \textit{lower bound} the objective of \eqref{eq:2STAGE-3}
by the quantity $\lambda_{max}\left(\mathbb{E}\left\{ \left.{\bf B}\left({\bf p}\left(t\right),t\right)\right|\mathscr{C}\left({\cal T}_{t-1}\right)\right\} \right)$,
resulting in the \textit{lower bound relaxation} 
\begin{equation}
\underset{{\bf p}\left(t\right)\in{\cal C}\left({\bf p}\left(t-1\right)\right)}{\mathrm{maximize}}\lambda_{max}\left(\mathbb{E}\left\{ \left.{\bf B}\left({\bf p}\left(t\right),t\right)\right|\mathscr{C}\left({\cal T}_{t-1}\right)\right\} \right).\label{eq:2STAGE-4}
\end{equation}
Apparently, the challenge now is to properly express the conditional
expectation involved as an explicit functional of ${\bf p}\left(t\right)$.
Interestingly, it can be shown that the random matrix $\boldsymbol{E}\left({\bf p}\left(t\right)\right)\triangleq\mathbb{E}\left\{ \left.{\bf B}\left({\bf p}\left(t\right),t\right)\right|\mathscr{C}\left({\cal T}_{t-1}\right)\right\} $
(where we emphasize the dependence on ${\bf p}\left(t\right)$) is
\textit{diagonal}, with elements 
\begin{align}
\boldsymbol{E}_{i} & \triangleq\mathbb{E}\left\{ \left.\dfrac{P_{0}\left|f_{i}\right|^{2}\left|g_{i}\right|^{2}-\zeta\sigma^{2}\left|g_{i}\right|^{2}}{P_{0}\left|f_{i}\right|^{2}+\sigma^{2}}\right|\mathscr{C}\left({\cal T}_{t-1}\right)\right\} ,\;i\in\mathbb{N}_{R}^{+}.\label{eq:Diagonal}
\end{align}
In order to evaluate the conditional expectations in each diagonal
element of $\boldsymbol{E}$, hereafter, we will assume that $1/\left(P_{0}\left|f_{i}\right|^{2}+\sigma^{2}\right)$
$\approx1/\left(P_{0}\left|f_{i}\right|^{2}\right)$, for all $i\in\mathbb{N}_{R}^{+}$,
corresponding to a \textit{high-SNR scenario} at the relays. This
approximation will be valid if either the noise power at the relays
is small, or the broadcasting power of the source is relatively large.
This situation is reasonable in beamforming networks, since the most
suitable network nodes in terms of information relaying may be selected
through a relay selection procedure. Then, \eqref{eq:Diagonal} becomes
\begin{equation}
\boldsymbol{E}_{i}\hspace{-2pt}=\mathbb{E}\left\{ \hspace{-2pt}\left.\left|g_{i}\right|^{2}\right|\mathscr{C}\left({\cal T}_{t-1}\right)\hspace{-2pt}\right\} \hspace{-2pt}-\dfrac{\zeta\sigma^{2}}{P_{0}}\mathbb{E}\left\{ \hspace{-2pt}\left.\left|g_{i}\right|^{2}\left|f_{i}\right|^{-2}\right|\mathscr{C}\left({\cal T}_{t-1}\right)\hspace{-2pt}\right\} ,\label{eq:Diagonal_NEXT}
\end{equation}
for all $i\in\mathbb{N}_{R}^{+}$. In fact, each $\boldsymbol{E}_{i}$
may be computed in closed form, as the following result suggests.
\begin{thm}
\textbf{\textup{(Conditional Correlations)}} Under the wireless channel
modeling assumptions of Section 3, it is true that
\begin{multline}
\mathbb{E}\left\{ \hspace{-2pt}\left.\left|g_{i}\right|^{2}\right|\mathscr{C}\left({\cal T}_{t-1}\right)\hspace{-2pt}\right\} \hspace{-2pt}\equiv\hspace{-2pt}10^{\rho/10}\hspace{-2pt}\exp\left(\vphantom{\dfrac{\log^{2}\left(10\right)}{200}\sigma_{\left.t\right|t-1}^{2,G_{i}}\left({\bf p}_{i}\left(t\right)\right)}\hspace{-2pt}\dfrac{\log\left(10\right)}{10}\mu_{\left.t\right|t-1}^{G_{i}}\hspace{-2pt}\left({\bf p}_{i}\left(t\right)\right)\hspace{-2pt}\right.\\
\left.+\dfrac{\log^{2}\left(10\right)}{200}\sigma_{\left.t\right|t-1}^{2,G_{i}}\hspace{-2pt}\left({\bf p}_{i}\left(t\right)\right)\right)\hspace{-2pt},\quad\text{and}\\
\mathbb{E}\left\{ \hspace{-2pt}\left.\left|g_{i}\right|^{2}\hspace{-2pt}\left|f_{i}\right|^{-2}\right|\mathscr{C}\left({\cal T}_{t-1}\right)\hspace{-2pt}\right\} \hspace{-2pt}\equiv\hspace{-2pt}\exp\hspace{-2pt}\left(\vphantom{\dfrac{\log^{2}\left(10\right)}{200}\begin{bmatrix}1\\
-1
\end{bmatrix}^{\boldsymbol{T}}\boldsymbol{\Sigma}_{\left.t\right|t-1}^{G_{i},F_{i}}\left({\bf p}_{i}\left(t\right)\right)\begin{bmatrix}1\\
-1
\end{bmatrix}}\hspace{-2pt}\dfrac{\log\left(10\right)}{10}\hspace{-2pt}\left(\mu_{\left.t\right|t-1}^{G_{i}}\hspace{-2pt}\left({\bf p}_{i}\left(t\right)\right)\right.\right.\\
\hspace{-2pt}\hspace{-2pt}\hspace{-2pt}\left.\left.\hspace{-2pt}-\mu_{\left.t\right|t-1}^{F_{i}}\hspace{-2pt}\left({\bf p}_{i}\left(t\right)\right)\right)\hspace{-2pt}+\hspace{-2pt}\dfrac{\log^{2}\left(10\right)}{200}\hspace{-2pt}\hspace{-2pt}\begin{bmatrix}1\\
-1
\end{bmatrix}^{\boldsymbol{T}}\hspace{-2pt}\hspace{-2pt}\boldsymbol{\Sigma}_{\left.t\right|t-1}^{G_{i},F_{i}}\hspace{-2pt}\left({\bf p}_{i}\left(t\right)\right)\hspace{-2pt}\begin{bmatrix}1\\
-1
\end{bmatrix}\hspace{-2pt}\right)\hspace{-2pt}.\label{eq:BOX_1}
\end{multline}
In \eqref{eq:BOX_1}, we define $\rho\triangleq10\mathbb{E}\left\{ \log_{10}\left|f_{i}^{MF}\left(t\right)\right|^{2}\right\} $,
and
\begin{flalign}
\mu_{\left.t\right|t-1}^{G_{i}}\left(\cdot\right) & \hspace{-2pt}\triangleq\hspace{-2pt}\alpha_{D}\left(\cdot\right)\ell+\boldsymbol{c}_{1:t-1}^{G_{i}}\left(\cdot\right)\boldsymbol{\Sigma}_{1:t-1}^{-1}\hspace{-2pt}\left(\boldsymbol{m}_{1:t-1}\hspace{-2pt}-\hspace{-2pt}\boldsymbol{\mu}_{1:t-1}\right),\\
\sigma_{\left.t\right|t-1}^{2,G_{i}}\left(\cdot\right) & \hspace{-2pt}\triangleq\hspace{-2pt}\sigma_{\xi}^{2}+\eta^{2}-\boldsymbol{c}_{1:t-1}^{G_{i}}\left(\cdot\right)\boldsymbol{\Sigma}_{1:t-1}^{-1}\left[\boldsymbol{c}_{1:t-1}^{G_{i}}\left(\cdot\right)\right]^{\boldsymbol{T}}\hspace{-2pt},\hspace{-2pt}\text{ with}\\
\boldsymbol{c}_{1:t-1}^{G_{i}}\left(\cdot\right) & \hspace{-2pt}\triangleq\hspace{-2pt}\left[\left\{ \boldsymbol{c}_{k}^{G_{i}}\right\} _{k\in\mathbb{N}_{t-1}^{+}}\right],
\end{flalign}
and 
\begin{flalign}
\boldsymbol{m}_{1:t-1} & \hspace{-2pt}\triangleq\hspace{-2pt}\left[\left\{ \boldsymbol{F}^{\boldsymbol{T}}\left({\bf p}\left(j\right),j\right)\,\boldsymbol{G}^{\boldsymbol{T}}\left({\bf p}\left(j\right),j\right)\right\} _{j\in\mathbb{N}_{t-1}^{+}}\right]^{\boldsymbol{T}},\\
\boldsymbol{\mu}_{1:t-1} & \hspace{-2pt}\triangleq\hspace{-2pt}\left[\left\{ \boldsymbol{\alpha}_{S}\left(j\right)\,\boldsymbol{\alpha}_{D}\left(j\right)\right\} _{j\in\mathbb{N}_{t-1}^{+}}\right]^{\boldsymbol{T}}\ell,\\
\boldsymbol{\Sigma}_{1:t-1} & \hspace{-2pt}\triangleq\hspace{-2pt}\begin{bmatrix}\boldsymbol{\Sigma}\left(1,1\right) & \cdots & \boldsymbol{\Sigma}\left(1,t-1\right)\\
\vdots & \ddots & \vdots\\
\boldsymbol{\Sigma}\left(t-1,1\right) & \cdots & \boldsymbol{\Sigma}\left(t-1,t-1\right)
\end{bmatrix},\\
\boldsymbol{c}_{k}^{G_{i}} & \hspace{-2pt}\triangleq\hspace{-2pt}\left[\left\{ \mathbb{E}\left\{ \sigma_{D}^{i}\left(t\right)\sigma_{S}^{j}\left(k\right)\right\} \hspace{-2pt}\right\} _{j}\,\left\{ \mathbb{E}\left\{ \sigma_{D}^{i}\left(t\right)\sigma_{D}^{j}\left(k\right)\hspace{-2pt}\right\} \hspace{-2pt}\right\} _{j}\right]\hspace{-2pt},
\end{flalign}
where $\boldsymbol{c}_{1:t-1}^{G_{i}}\left(\cdot\right)\in\mathbb{R}^{1\times\left(t-1\right)2R}$,$\boldsymbol{m}_{1:t-1}$
and $\boldsymbol{\mu}_{1:t-1}$ are in $\mathbb{R}^{\left(t-1\right)2R\times1}$
and $\boldsymbol{\Sigma}_{1:t-1}$ is in $\mathbb{S}_{++}^{\left(t-1\right)2R}$,
for all $t\in\mathbb{N}_{N_{T}}^{+}$. In \eqref{eq:BOX_1}, we have
\begin{flalign}
\boldsymbol{\Sigma}_{\left.t\right|t-1}^{G_{i},F_{i}}\left(\cdot\right) & \equiv\hspace{-1.505pt}\begin{bmatrix}\eta^{2}+\sigma_{\xi}^{2} & \eta^{2}e^{-\frac{\left\Vert {\bf p}_{S}-{\bf p}_{D}\right\Vert _{2}}{\delta}}\\
\eta^{2}e^{-\frac{\left\Vert {\bf p}_{S}-{\bf p}_{D}\right\Vert _{2}}{\delta}} & \eta^{2}+\sigma_{\xi}^{2}
\end{bmatrix}\nonumber \\
 & -\begin{bmatrix}\boldsymbol{c}_{1:t-1}^{G_{i}}\left(\cdot\right)\\
\boldsymbol{c}_{1:t-1}^{F_{i}}\left(\cdot\right)
\end{bmatrix}\boldsymbol{\Sigma}_{1:t-1}^{-1}\begin{bmatrix}\boldsymbol{c}_{1:t-1}^{G_{i}}\left(\cdot\right)\\
\boldsymbol{c}_{1:t-1}^{F_{i}}\left(\cdot\right)
\end{bmatrix}^{\boldsymbol{T}}\in\mathbb{R}^{2\times2},
\end{flalign}
and the quantities $\mu_{\left.t\right|t-1}^{F_{i}}\left(\cdot\right)$
and $\boldsymbol{c}_{1:t-1}^{F_{i}}\left(\cdot\right)$ are defined
accordingly, swapping ``$D$'' and ``$S$'' where applicable.
\end{thm}
Theorem 1 implies that \eqref{eq:Diagonal_NEXT} can be explicitly
expressed in terms of the available CSI and the related statistics.
This is possible due to the Gaussianity of the log-squared magnitude
of the observed channels. The proof involves the evaluation of the
related conditional moment generating functions and it is omitted
due to lack of space. 

From \eqref{eq:BOX_1}, one can observe that \eqref{eq:Diagonal_NEXT}
is a well defined functional of ${\bf p}_{i}\left(t\right)$, that
is, $\boldsymbol{E}_{i}\equiv\boldsymbol{E}_{i}\left({\bf p}_{i}\left(t\right)\right)$,
for all $i\in\mathbb{N}_{R}^{+}$. Additionally, as it was expected,
each $\boldsymbol{E}_{i}$ is \textit{independent of all} ${\bf p}_{j}\left(t\right),j\neq i$.
Under these circumstances, the program under consideration, described
by \eqref{eq:2STAGE-4}, can be expressed as the maximization of $\boldsymbol{E}_{i}\left({\bf p}_{i}\left(t\right)\right)$
jointly over ${\bf p}\left(t\right)\in{\cal C}\left({\bf p}\left(t-1\right)\right)$
and $i\in\mathbb{N}_{R}^{+}$, which is equivalent to
\begin{equation}
\underset{i\in\mathbb{N}_{R}^{+}}{\mathrm{max}}\left[\underset{{\bf p}_{i}\left(t\right)\in{\cal C}\left({\bf p}_{i}\left(t-1\right)\right)}{\mathrm{max}}\boldsymbol{E}_{i}\left({\bf p}_{i}\left(t\right)\right)\right].\label{eq:2STAGE-5}
\end{equation}
In particular, each of the \textit{$R$ two dimensional} problems
involved can be performed \textit{locally at each relay}, provided
the availability of common global information (channel magnitudes
and relay positions).

Quite remarkably, the discussion above reveals that the optimal control
policy is of a purely selective form: At each $t\in\mathbb{N}_{N_{T}}^{+}$
and provided that the inner problem of \eqref{eq:2STAGE-5} can be
solved exactly, only the $i^{o}$-th relay should move, where $i^{o}$
is the respective solution of the outer maximization of \eqref{eq:2STAGE-5};
in fact, either any of the rest of the relays moves or not is irrelevant.

\vspace{-0.2cm}

\subsection{Determination of Relay Motion Controllers}

What remains now is to determine the controllers of the $i^{o}$-th
relay, selected to carry out the decision that optimizes the relaxed
cost of \eqref{eq:2STAGE-4}. Suppose that ${\bf p}^{o}\left(t\right)\in\underset{{\bf p}_{i}\left(t\right),i}{\mathrm{arg\,max}}\boldsymbol{E}_{i}\left({\bf p}_{i}\left(t\right)\right)$
has been determined, for instance, numerically. Then, it suffices
to fix a path in ${\cal S}$, such that the points ${\bf p}^{o}\left(t\right)$
and ${\bf p}_{i^{o}}\left(t-1\right)$ are connected in at most time
$\Delta\tau_{t}$. By far the easiest choice for such a path is the
straight line connecting ${\bf p}^{o}\left(t\right)\triangleq\left[x_{t}^{o}\,y_{t}^{o}\right]^{\boldsymbol{T}}$
and ${\bf p}_{i^{o}}\left(t-1\right)\triangleq\left[x_{t-1}^{i^{o}}\,y_{t-1}^{i^{o}}\right]^{\boldsymbol{T}}$.
Therefore, we can choose the optimal relay motion controller as
\begin{equation}
{\bf u}_{t-1}^{o}\left(\tau\right)\triangleq\dfrac{1}{\Delta\tau_{t-1}}\begin{bmatrix}y_{t}^{o}-y_{t-1}^{i^{o}}\\
x_{t}^{o}-x_{t-1}^{i^{o}}
\end{bmatrix},\quad\tau\in\Delta\tau_{t-1},
\end{equation}
completing the presentation of the proposed solution to the mobile
beamforming problem under consideration.

\vspace{-0.2cm}

\section{Conclusion}

\vspace{-0.1cm}

We have considered the problem of stochastic relay spatial control
for beamforming optimization in single-source single-destination robotic
relay networks. Under a realistic spatiotemporal stochastic model
for the communication medium, we proposed a $2$-stage stochastic
programming formulation for specifying relay spatial controllers,
such that the future \textit{expected} \textit{reciprocal} of their
total beamforming power is maximized, based only on causal CSI at
the relays. Due to the intractability of the original problem, we
have proposed a lower bound relaxation, which is equivalent to a set
of tractable two dimensional subproblems, solved at each relay independently.
Interestingly, the aforementioned formulation results to a relay selection
plus control scheme; at each time slot, only one relay should move
- the one resulting to the highest expected beamforming improvement.
This work essentially serves as a basis for several extended formulations
of the mobile beamforming problem and more generally of related problems
in spatially controlled communication systems; these constitute the
subject of current research.

\newpage{}

\bibliographystyle{IEEEbib}
\bibliography{IEEEabrv}

\end{document}